# Thermal emissions and climate change: Cooler options for future energy technology

Nick E.B. Cowern[1,2] and Chihak Ahn[1]

[1]*School of Electrical, Electronic and Computer Engineering, Newcastle University, Newcastle upon Tyne, NE1 7RU, UK*

[2]*Institute for Research on Environment and Sustainability, Newcastle University, Newcastle upon Tyne, NE1 7RU, UK*

**Summary**

Climate projections that account for both greenhouse gas emissions and human-made *thermal* emissions show that global temperature forcing will not peak and decline as currently expected, but will continue to rise indefinitely, unless we exploit renewable energy sources that are heat-neutral or act as heat sinks.

**Abstract**

Global warming arises from 'temperature forcing', a net imbalance between energy fluxes entering and leaving the climate system and arising within it. Humanity introduces temperature forcing through greenhouse gas emissions, agriculture, and thermal emissions from fuel burning. Up to now climate projections, neglecting thermal emissions, typically foresee maximum forcing around the year 2050, followed by a decline. In this paper we show that, if humanity's energy use grows at 1%/year, slower than in recent history, and if thermal emissions are not controlled through novel energy technology, temperature forcing will increase indefinitely unless combated by geoengineering. Alternatively, and more elegantly, humanity may use renewable sources such as wind, wave, tidal, ocean thermal, and solar energy that exploit energy flows already present in the climate system, or act as effective sinks for thermal energy.

Despite decades of significant global warming, humanity is only now beginning significantly to address the reduction of $CO_2$ emissions caused by power generation and transport (*1*). It is now clear that $CO_2$ emissions must be largely eliminated during the first half of this century in order to minimise the risk of dangerous climate change (*2,3,4*). However, what has not been widely understood is the likely climate impact of thermal emissions from power generation and use, which may cause significant additional warming beyond the middle of this century.

Energy technologies such as nuclear (fission or fusion), fossil fuels and geothermal power plants are human-made sources of heat energy which flows into Earth's climate system. Such thermal emissions contribute directly to Earth's heat budget and cause global warming. In contrast, most renewable energy technologies, such as wind, wave and tidal power, harvest energy from Earth's dissipative systems, and thus do not directly add to Earth's heat budget. Solar electricity generation, a promising and fast-expanding energy technology, acts in a more complex way because it exploits an existing energy flow (incoming solar radiation) but in so doing, for the purpose of efficient energy generation, it typically lowers the albedo of Earth's surface at the solar collector location, thus adding to Earth's heat budget. Still, the thermal impact of solar generation may be less than that of heat-based energy sources like nuclear and geothermal power, because solar collectors take the place of terrain which was already absorbing a significant proportion, typically from 60–90%, of incident solar energy.

The flow of human-made heat into the climate system plays only a small part in present-day global warming, but as the world moves to a low-carbon energy economy increasingly dominated by electricity generation, this transition, together with expected growth in consumption, will lead to serious warming effects in addition to those previously caused by human-made $CO_2$. At present, the reduction of $CO_2$ emissions must be humanity's paramount concern, and any cost-effective zero-carbon technology is preferable to a carbon emitting one. However by midcentury technologies will need to be in place to generate usable energy without significant thermal emissions integrated over the full cycle of generation, transmission and energy consumption. This consideration has major implications for long-range funding choices between competing energy technologies such as fusion, wind, and solar energy, which could potentially contribute substantial proportions of the world's energy supply from midcentury onwards.

In this paper we begin by considering the global temperature forcing arising from thermal emissions from heat-based energy sources - fuel burning and geothermal power. The resultant forcing is compared to a typical estimate of $CO_2$ forcing assuming responsible measures are taken to control $CO_2$ emissions (*2*). Thermal emissions are shown to contribute increasingly to total forcing, threatening to prevent the decline in forcing from midcentury onwards which climate scientists have assumed will occur after $CO_2$ emissions have fallen significantly (*1*). We then turn to an evaluation of the likely impact of solar energy, considering various scenarios for collector albedo based on different types of solar technology. One of these options could, speculatively, lead to solar power generation combined with a net negative temperature forcing. Finally, we consider the impact of ocean thermal energy conversion, which contributes a transient, but potentially large, negative temperature forcing.

Current global primary energy use is increasing at about 2%/yr (*5*), and apart from short-term variations is likely to continue increasing for the foreseeable future. Following the assumptions of Ref. 6 we assume a constant growth rate of 1%/yr with a transition to a zero-carbon energy economy based on electrical generation, occurring during the period up to 2100. As a baseline for considering a transition to renewable technologies, we first evaluate a scenario where this transition is based on nuclear and/or fossil fuels with carbon capture and storage, assuming an electrical generation efficiency of 35-50% (*7*).

The resulting thermal forcing in $W/m^2$ is plotted as the pale red band in Fig. 1, together with the $CO_2$ forcing (black curve) resulting from emissions in the 'Coal Phase-Out' scenario of Kharecha and Hansen (*2*). Instead of peaking and subsequently decreasing from midcentury onwards as in Ref. (*2*), the total forcing from $CO_2$ and thermal emissions stabilises for about 100 years at a level of nearly 3 $W/m^2$, corresponding to an equilibrium temperature rise of about 3 - 4°C (*3*), and then the forcing rises further (full red band). Based on virtually all accepted climate models, this would lead Earth into a period of dangerous climate change, either late this century or early during the next one.

If heat-based energy sources could be fully supplanted by renewables such as wind, wave or solar energy, thermal emissions would be much less significant – only heat generated in plant construction and maintenance, and possibly second-order changes in the climate system owing to perturbations of natural energy flows by these energy conversion systems, would play a role. However, current forecasts suggest that such energy sources, while important, cannot supply all of humanity's energy needs, and much research, technology development and manufacturing is currently being devoted to solar-based electricity generation, with the photovoltaic (PV) market growing at a near-term projected rate of nearly 50% according to some estimates (*8*).

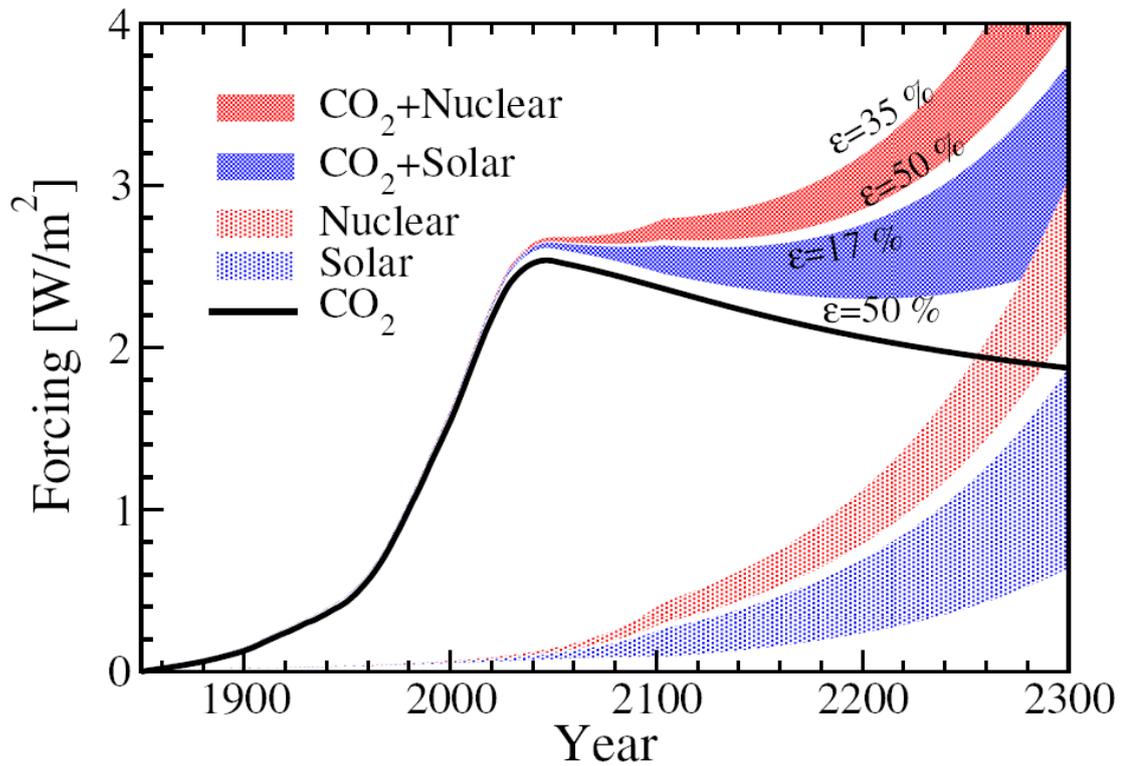

**Fig. 1:** Global temperature forcings due to human-made $CO_2$ and thermal effects. The $CO_2$ forcing is based on the 'Coal Phase-out' scenario of Kharecha and Hansen, and the thermal contribution ('Nuclear' or 'Solar') is based on 1%/yr growth of total energy use, with electrical generation efficiency in the indicated range, and non-electrical energy (phased out between 2020 to 2100) approximated as 100% efficient.

In a recently published Solar Grand Plan, Zweibel *et al.* have proposed a strategy to transform the US energy economy from its current fossil-fuel rich mix to one dominated by solar power (*6*). In their scenario fuel costs are kept at acceptable values, and total energy production grows by 1%/yr during the course of this century. By 2100 the transition to a solar energy economy with modest contributions from wind and geothermal power is essentially complete as solar electrical output plus other renewables reach over 90% of the total energy supply including transport. Zweibel et al. (*6*) foresee a solar collection area of 400,000 km$^2$ in the U.S. by 2100, which we extrapolate (*9*) to a worldwide collection area of $1.8 \times 10^6$ km$^2$ (about 0.3% of Earth's surface) – a 'Global Solar Grand Plan'. Solar energy implementation at this scale could produce substantial climate impacts.

Fig. 2 shows schematically the impact of a solar PV-based power plant on local and global energy balance, taking into account the transmission of electrical energy away from the collector area towards distant consumers. Installation of the PV collector array increases local solar absorption because the terrain albedo, $a_t$, typically 0.25 – 0.45 for a hot, flat desert environment (*10,11*), is reduced to the effective albedo of the PV array, $a_c$. In most current PV technologies, $a_c \approx 0$, since effective reflectivities of solar collector surface films are currently in the range of a few per cent and falling as further technical advances are made. The net increase in solar flux absorbed within the collector area, $A$, is then $\delta\phi = (a_t - a_c)\phi$, where $\phi$ is the incident solar flux. The amount of this energy flux transmitted as electricity to load, $\varepsilon\phi$, is dependent on the solar conversion efficiency of the installed collector plant, $\varepsilon$, however at remote locations virtually all of this transmitted energy ultimately converts to heat, thus the net increase in global heat input per second, $\Delta W_G$, is simply

$$\Delta W_G = (a_t - a_c)\phi A, \qquad (1)$$

where A is the global area covered by solar collectors. We now address the issue of global temperature forcing represented by this equation.

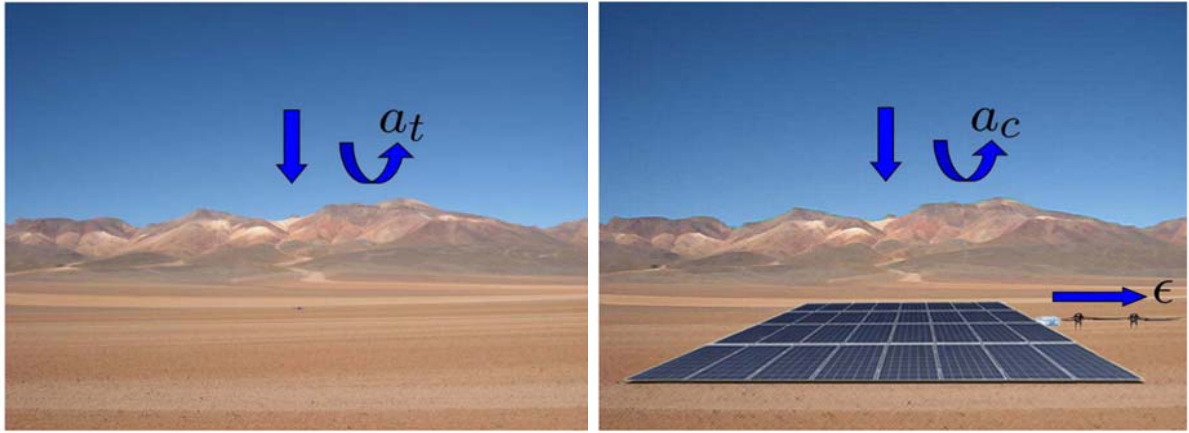

**Fig. 2:** Schematic view of a solar PV-based power plant before and after installation. The albedo of the rectangular area changes from $a_t$, that of the local terrain, to $a_c$, the albedo of the PV array, leading to an change in the energy flux entering this area. Local climate is further modified by the outflow of electrical energy to the transmission network or storage site, however this relocation of energy does not alter net global temperature forcing since virtually all the relocated energy ultimately transforms into heat.

In order to do this we again assume 1%/yr growth of total energy use, and within this, following Zweibel *et al.*, the solar contribution is assumed to grow by about 6%/yr between 2020 and 2100, after which it becomes the dominant energy source and grows by 1%/yr constrained by our assumption on total energy use. Excess heat production is calculated assuming an average original terrain albedo of 0.3 and solar efficiencies in the range 17 - 50%. All other solar energy incident on the collector array is assumed to be absorbed and thus contribute to heating. The resulting global temperature forcing is plotted as the blue-shaded band for different values of $\varepsilon$, in Fig. 1. Qualitatively the forcing is similar to that caused by nuclear energy, but the effect is smaller and decreases as solar conversion efficiency is increased. The improvement with respect to heat-based energy generation arises from the factor $a_t - a_c$ in equation (1); the excess heat produced by solar power is equal to the radiant energy absorbed by the collector, *minus* that which the original terrain would have absorbed as part of Earth's normal radiation balance.

One way to compensate for the temperature forcing caused by solar power is to use a form of geoengineering (*12*) known as 'albedo engineering' in which an area of relatively low albedo is replaced by a high-albedo surface. For the highest solar efficiency of 50%, the area of high-albedo surface needed to compensate this is roughly an order of magnitude smaller than the solar collection area, depending on the terrain albedo where it is installed. Thus, if solar collectors are economically feasible, this additional technology is presumably feasible as well.

Another, more speculative, approach incorporates albedo engineering directly into PV technology by backing a thin active PV layer with a reflective, or partially reflective, substrate. Given a suitable choice of PV material and device structure, this enables much of the unused energy in the long-wavelength spectrum of sunlight (below the semiconducting band gap) to be reflected back out of the entry surface of the PV cell, thus raising its effective albedo. While a full evaluation of this effect is beyond the scope of this paper, we make some simplified model calculations. We consider an idealised PV cell with a simple active layer 1 (coloured blue) and a substrate layer 2 (coloured red) as illustrated in Fig. 3a. Photons with energy $E$ above the bandgap $E_g$ of layer 1 are assumed to be converted with 100% quantum efficiency, i.e. with energy efficiency $E_g/E$, while a proportion $A$ of photons below the bandgap energy are assumed to be absorbed in layer 2, i.e. not reflected back through the entry surface or transmitted through layer 2. For simplicity we assume that A is independent of wavelength. The quantities R and T shown in Fig. 3a are the fractions of the sub-bandgap light reflected and transmitted, respectively, by layer 2.

In this simple model the energy absorbed by the cell is a sum of the energy converted to charge carriers in layer 1 and the energy absorbed in layer 2, i.e. $N_> \overline{E}_> + A N_< \overline{E}_<$, where $N$ indicates the photon flux and $\overline{E}$ the mean photon energy in the relevant energy range, above (>) or below (<) the bandgap.

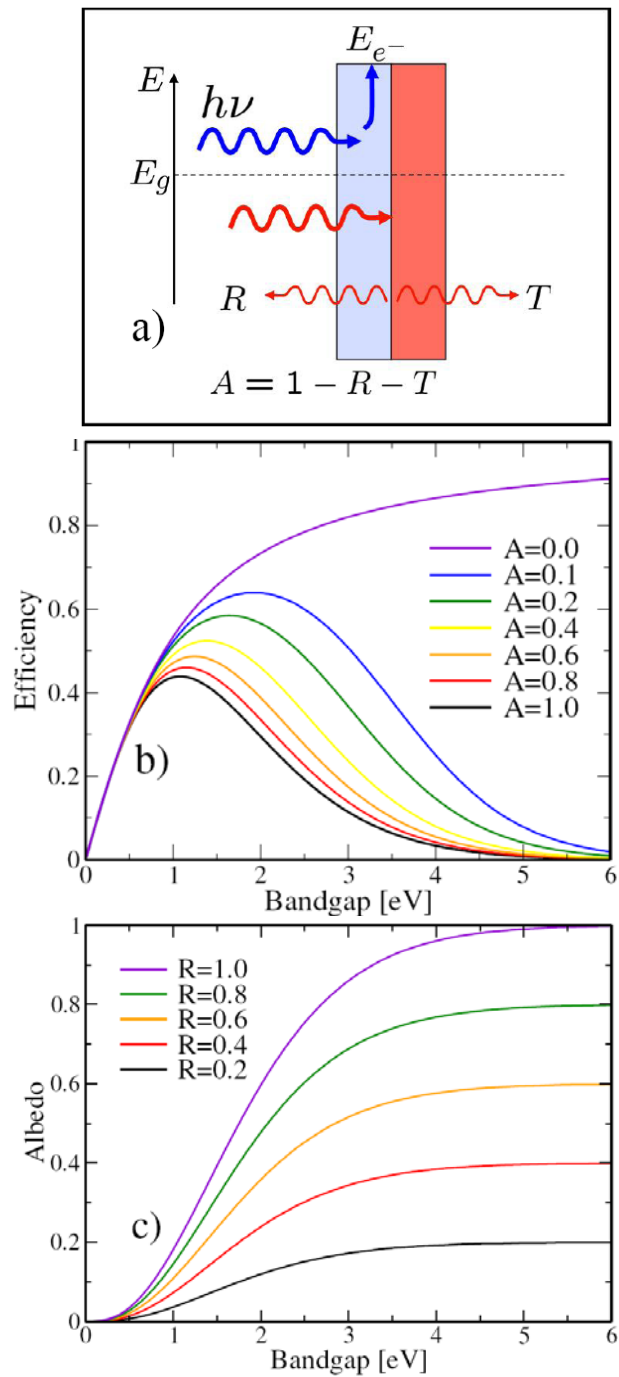

**Fig. 3:** Radiation fluxes and photoelectron generation in a simplified one-dimensional PV cell model. a) - model cell schematic. b) - thermal efficiency in the model PV cell. c) - approximate effective albedo of the model cell. The combination of high thermal efficiency and high albedo at bandgaps in the 2-eV range illustrates the potential of such wide bandgap PVs for large-scale solar 'cool power' generation.

The 'thermal efficiency' of our model cell, defined as the fraction of this absorbed energy converted to electrical energy, is

$$\varepsilon = \frac{E_g}{\overline{E}_> + (N_< / N_>) A \overline{E}_<} \quad (2)$$

In general this efficiency is higher than that defined in terms of electrical output divided by incident solar energy flux, because the denominator includes only the absorbed light. For a reflective solar cell, i.e. one with no transmission through the back of the cell, $A=1-R$ and the effective albedo of the cell is given by the ratio of reflected to incident energy fluxes,

$$a_c = \frac{R N_< \overline{E}_<}{N_> \overline{E}_> + N_< \overline{E}_<} \quad (3)$$

Fig. 3b shows the thermal efficiencies of cells with values of $A$ in the range from 0 to 1, and Fig. 3c shows the corresponding effective albedo values. For bandgap values ranging from that of crystalline Si (1.2 eV) up to 2.5 eV, and for A=0.1, thermal efficiency is in the range ~55-65%, higher than can be achieved with conventional energy technologies. In addition, cell albedo is typically higher than terrain albedo, offering a theoretical possibility that PV technology could produce a negative temperature forcing supporting global cooling.

As an example, we present the time evolution of temperature forcing for a global solar grand plan using reflective PV technology with R=0.9 (A=0.1). The impact of such a technology is shown in Fig. 4, the higher value (red dashed curve) corresponding to the silicon bandgap and the lower values (orange and green dashed curves) corresponding to wide-bandgap PV materials with $E_g = 2$ eV and 2.5 eV, respectively.

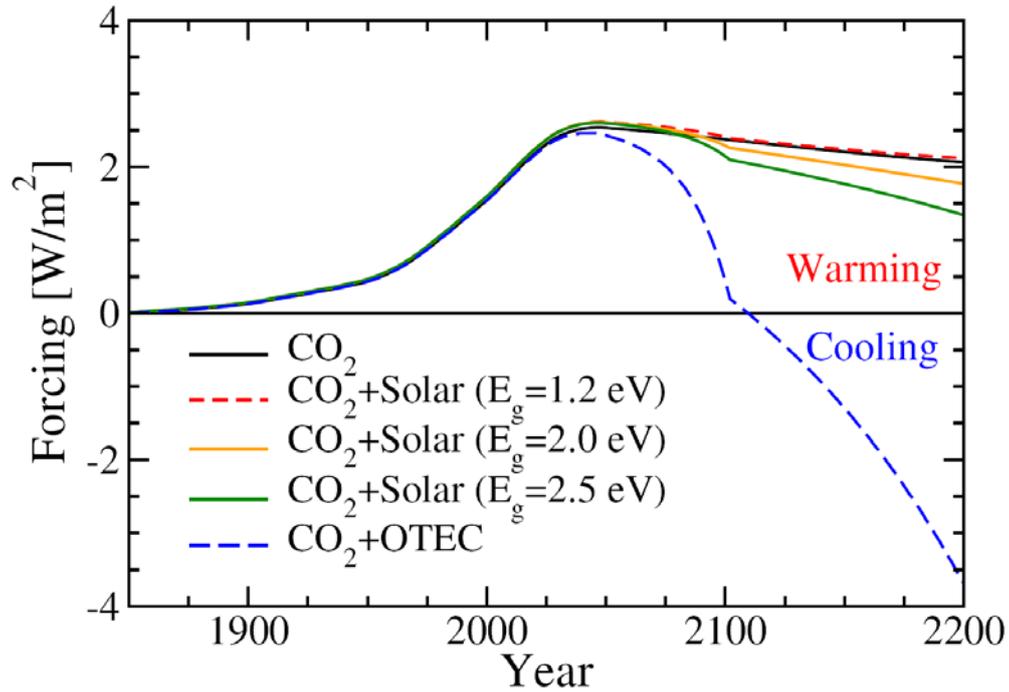

**Fig. 4:** Global temperature forcing versus time as a result of $CO_2$ emissions (black solid curve), $CO_2$ emissions plus thermal effects from a global solar grand plan using reflective silicon PVs with a bandgap of 1.2 eV (red dashed curve), and the results for $CO_2$ emission plus PVs with wider band-gaps of 2 and 2.5 eV (orange and green solid curves). In the wide band-gap cases the positive forcing from $CO_2$ emissions is significantly reduced. Finally, we show the impact in the idealized case where all global energy is generated using ocean thermal energy conversion. In this case, temperature forcing turns negative by 2100 (blue dashed curve), most probably returning global temperature to historical levels during the next century.

The curve for net global temperature forcing using the wider bandgaps now lies *below* the prediction for $CO_2$ forcing because of the cooling effect of the wide bandgap PV technology. Although the trend of these results is clear, the predicted curves for the different PV materials in Fig. 4 are not quantitatively exact, given the simplifications used in our analysis.

Finally, we show the potential impact of the developing technology of ocean thermal energy conversion (OTEC), which generates electricity by pumping heat from warm ocean-surface waters to the cooler, deeper ocean (*13*). Here the heat sunk per output electrical energy, $1/\varepsilon$, is high due to the relatively low thermodynamic conversion efficiency, $\varepsilon$, of heat pump technology. This determines the amount of heat taken from the climate system after accounting for use of electrical output power,

$$-\Delta W_G = P(\varepsilon^{-1} - 1)$$

where *P* is global electrical output and the term $-1$ accounts for heat created from electricity use. We assume as in most climate models that the ocean surface couples strongly (i.e. rapidly) to the atmosphere, we treat the slow transport of buried heat back to the ocean surface as negligible on the time scale of our predictions, and we assume $\varepsilon = 0.06$, close to the maximum expected from OTEC model simulations (*14*), so providing a conservative estimate of $\varepsilon^{-1}$.

The result, shown in Fig. 4 by the blue dashed curve, is dramatic. Even a substantially smaller contribution of OTEC to global energy generation, producing a proportionately smaller negative temperature forcing, could be an important contribution to stabilising global surface temperature. The key principle here is that heat is pumped to the deep ocean much faster than is achieved by natural ocean heat transport processes. In this way, OTEC, with an appropriate magnitude and spatial distribution of generating capacity, could help control and even reverse the rising trend of ocean surface temperature which is driving fast, potentially dangerous, climate feedbacks. These ideas appear, in outline, to offer a synergistic combination of power generation and environmentally compliant geoengineering for responsible future energy use.

We have shown that thermal effects from human energy consumption will play an increasingly significant role in global temperature forcing in the future. Consequently it is important to discriminate between renewable energy sources that inject heat into Earth's climate system (geothermal energy), those that rely on Earth's dissipative systems (wind, wave, tidal energy), and those that may potentially remove heat energy (suitably chosen solar technology, OTEC, and perhaps other future technologies). Correct technology choices will reduce the magnitude and time period of future global warming caused by current $CO_2$ emissions. Conversely, nuclear fusion, which may potentially come on stream as a significant energy source several decades hence, will be too late as a replacement for $CO_2$-emitting technologies, and inherently (*15*) will not meet contemporaneous thermal emissions criteria for a sustainable global environment. We suggest a re-evaluation by the science and engineering communities, taking thermal cycle analysis into account, so that the most promising future technologies for zero-carbon, thermally-compliant energy generation can be targeted for research and development during the next decade.

1. IPCC Fourth Assessment Report – "Climate Change 2007: Synthesis Report", www.ipcc.ch/pdf/assessment-report/ar4/syr/ar4_syr.pdf

2. P.A. Kharecha, J.E, Hansen, *Global Biogeochem. Cycles* **22**, GB3012 (2008).

We thank N.S. Bennett for discussions and critical reading of the manuscript.